\documentclass[aps,prb,showpacs,twocolumn]{revtex4-1}
\usepackage{amssymb}
\usepackage{amsmath}
\usepackage{graphicx}
\usepackage{dsfont}
\usepackage{times}

\begin{document}

\title{Power-Law Random Banded Matrix Ensemble as the Effective Model for
Many-Body Localization Transition}
\author{Wen-Jia Rao}
\email{wjrao@hdu.edu.cn}
\affiliation{School of Science, Hangzhou Dianzi University, Hangzhou 310027, China.}
\date{\today }

\begin{abstract}
We employ the power-law random band matrix (PRBM) ensemble with single
tuning parameter $\mu $ as the effective model for many-body localization
(MBL) transition in random spin systems. We show the PRBM accurately
reproduce the eigenvalue statistics on the entire phase diagram through the
fittings of high-order spacing ratio distributions $P(r^{(n)})$ as well as
number variance $\Sigma ^{2}(l)$, in systems both with and without
time-reversal symmetry. For the properties of eigenvectors, it's shown the
entanglement entropy of PRBM displays an evolution from volume-law to
area-law behavior which signatures an ergodic-MBL transition, and the
critical exponent is found to be $\nu =0.83\pm 0.15$, close to the value
obtained in 1D physical model by exact diagonalization while the
computational cost here is much less.
\end{abstract}

\maketitle

\section{Introduction}

The phase of matter in isolated quantum systems has attracted a lot of
attention in the past decade, it is by now well-established the existence of
two generic phases: an ergodic phase and a many-body localized (MBL) phase%
\cite{Gornyi2005,Basko2006}. In an ergodic phase, the system acts as the
heat bath for its subsystem, hence the quantum entanglement follows
volume-law and exhibits ballistic spreading after quantum quench. In
contrast, a MBL phase is where localization persists in the presence of weak
interactions, which leads to area-law entanglement and slow entanglement
spreading. The different scaling behaviors of quantum entanglement provide
the modern understanding about these two phases\cite%
{Kjall2014,Yang2015,Serbyn16,Gray2017,Maksym2015,Kim,Bardarson,Abanin,Znidaric}%
.

More traditionally, the ergodic and MBL phase are distinguished by their
eigenvalue statistics\cite%
{Oganesyan,Avishai2002,Regnault16,Regnault162,Huse1,Huse2,Huse3,Garcia,Luitz}%
, whose mathematical foundation is laid by the random matrix theory\cite%
{Mehta,Haake}. The eigenvalues of an ergodic phase are well-correlated,
whose distribution belongs to the Wigner-Dyson (WD) class which further
divides into three ensembles depending on the system's symmetry: the
Gaussian orthogonal/unitary ensemble (GOE/GUE) for orthogonal/unitary
systems with/without time reversal symmetry, and Gaussian symplectic
ensemble (GSE) for time-reversal invariant systems with broken spin
rotational invariance. On the contrary, the eigenvalues in MBL phase are
independent of each other and belongs to the Poisson ensemble.

Compared to the properties of each phase, much less is known about the
evolution in between. Up to now, there have been a number of proposed random
matrix models accounting for the spectral statistics right at the critical
point, or even along the whole phase diagram\cite%
{Shukla0,Serbyn,SRPM,Sierant19,Rao21,Rao212,Buijsman,Sierant20}, among which
two outstanding models are the Gaussian $\beta $ ensemble\cite{Buijsman} and
the $\beta -h$ model\cite{Sierant20}. The former generalizes the WD classes
into the one with continuous Dyson index $\beta \in (0,\infty )$, and is
shown to faithfully reproduce the short-range level correlations for the MBL
system, but the deviations soon become non-negligible when long-range
correlations are considered. On the other hand, the two-parameter $\beta -h$
model is shown to accurately describe the level correlations on both short
and long ranges during MBL transition. However, this model is based on the
joint probability distribution of eigenvalues and lacks clear matrix
construction, hence cannot provide insights about the eigenvectors.

Given above facts, we impose three basic requirements for an optimal
effective model for an MBL system: (i) it should accurately reproduce the
eigenvalue statistics of the physical model -- on both short and long ranges
-- on the entire phase diagram; (ii) it should have well-defined matrix
construction, whose eigenvectors should further reflect the universal
properties of physical system, such as the scaling of entanglement; (iii) it
should contain as less tuning parameters as possible.

In this work, we employ the power-law random banded matrix (PRBM)\cite%
{Mirlin1} with fixed $B=1$ and single tuning parameter $\mu$ (definitions
come in next section) as the effective model for studying MBL transition.
PRBM was introduced as the critical random matrix model for Anderson
localization transition, which contains all the key features of the latter
such as multifractality of eigenvectors and spectrum compressiblity, as
verified in a considerate number of works\cite%
{Mirlin2,Mirlin3,Mirlin4,Varga2,Varga3,Varga4,Kravtsov1997,Varga,Cao}%
. And here, we bring PRBM to the many-body regime. Remarkably, we find
nearly perfect agreement for the eigenvalue statistics between the
single-parameter PRBM and typical physical models over the entire phase
diagram, in cases both with and without time-reversal symmetry. Moreover,
for the eigenvectors, we find PRBM holds both volume-law and area-law
behavior for its entanglement entropy (EE), which signatures an ergodic-MBL
transition, and the critical exponent is found to be $\nu=0.83\pm0.15$,
close to that obtained by exact diagonalization for physical model\cite{RD}
while the computational cost here is much less.

This paper is organized as follows. In Sec.\ref{sec2} we introduce the PRBM.
In Sec.\ref{sec3} we use PRBM to fit the eigenvalue statistics of typical
physical models, both with and without time-reversal symmetry. In Sec.\ref%
{sec4} we study the entanglement scaling in PRBM, and determine the
corresponding critical exponent. Discussion and conclusion come in Sec.\ref%
{sec5}.

\section{Power-Law Random Banded Matrix}

\label{sec2}

The power-law random banded matrix (PRBM) ensemble\cite{Mirlin1} is a
Gaussian ensemble of $N\times N$ dense Hermitian matrices $H$ with random
elements, whose distribution satisfy%
\begin{eqnarray}
\langle H_{ij}\rangle &=&0\text{, }\quad \langle \left( H_{ii}\right)
^{2}\rangle =\beta ^{-1}\text{,} \\
\langle \left( H_{ij}^{\alpha }\right) ^{2}\rangle &=&\beta ^{-1}\left[
1+\left( \left\vert j-i\right\vert /B\right) ^{2\mu }\right] ^{-1}\text{, }%
\alpha =1,2,..\beta  \notag
\end{eqnarray}%
where $\beta =1,2,4$ is the Dyson index, $H_{ij}^{\alpha }$ is the $\alpha $%
-th component of the non-diagonal element $H_{ij}$. For instance, in the
unitary PRBM with $\beta =2$, $H_{ij}^{1}$ and $H_{ij}^{2}$ are the real and
imaginary part of $H_{ij}$.\ PRBM is controled by the tuning parameter $\mu
\in \left( 0,\infty \right) $: in the limit $\mu \rightarrow 0$, we have $%
\langle \left( H_{ij}^{\alpha }\right) ^{2}\rangle \rightarrow 1$, hence $H$
becomes the standard random matrix in WD classes for the ergodic phases;
while in the limit $\mu \rightarrow \infty $, we have $\langle \left(
H_{ij}^{\alpha }\right) ^{2}\rangle \rightarrow 0$ for $i\neq j$, which
means all non-diagonal elements vanish and $H$ becomes a random diagonal
matrix for the Poisson ensemble. In cases with $B\gg 1$ and $B\ll 1$, the
critical point is analytically proved to be $\mu _{c}=1$, separating the
metallic ($\mu <1$) and localized ($\mu >1$) phase in the Anderson model\cite%
{Mirlin1,Mirlin2,Mirlin3}. In this work, we concentrate on the case with $%
B=1 $ which is by far unaccessible by analytical treatment and leave $\mu $
as the single varying parameter. The criticality of $\mu _{c}=1$ in this
case has been numerically verified in earlier works\cite{Varga,PRBMS}, and
will be further justified in this study.

The well-defined matrix construction of PRBM allows us to numerically study
both its eigenvalues and eigenvectors, and we start with the former. As the
first step, we show PRBM ($B=1$) with varying $\mu $ can describe the
interpolation between WD and Poisson. Specifically, we numerically
diagonalize $400$ samples of PRBM\ at various $\mu $s, with the matrix
dimension $N=2000$, and take $400$ eigenvalues in the middle to determine
eigenvalue statistics. For the latter, we study the distribution of spacing
ratios, whose definition is\cite{Oganesyan}%
\begin{equation}
r_{i}^{\left( n\right) }=\frac{E_{i+2n}-E_{i+n}}{E_{i+n}-E_{i}}\text{.}
\end{equation}%
The form of $P\left( r^{\left( n\right) }\right) $ with $n=1$ has been
analytically derived in Ref.~[\onlinecite{Atas}] and later generalized to
higher-order ones to incorporate level correlations on longer ranges\cite%
{Tekur,Rao20}. Compared to the more traditional quantities like level
spacings $\left\{ s_{i}=E_{i+1}-E_{i}\right\} $ or number variance $\Sigma
^{2}\left( l\right) $, spacing ratios are independent of density of states
and requires no unfolding procedure, which is non-unique and may raise
subtle misleading signatures in certain system\cite{Gomez2002}.

\begin{figure}[t]
\centering
\includegraphics[width=\columnwidth]{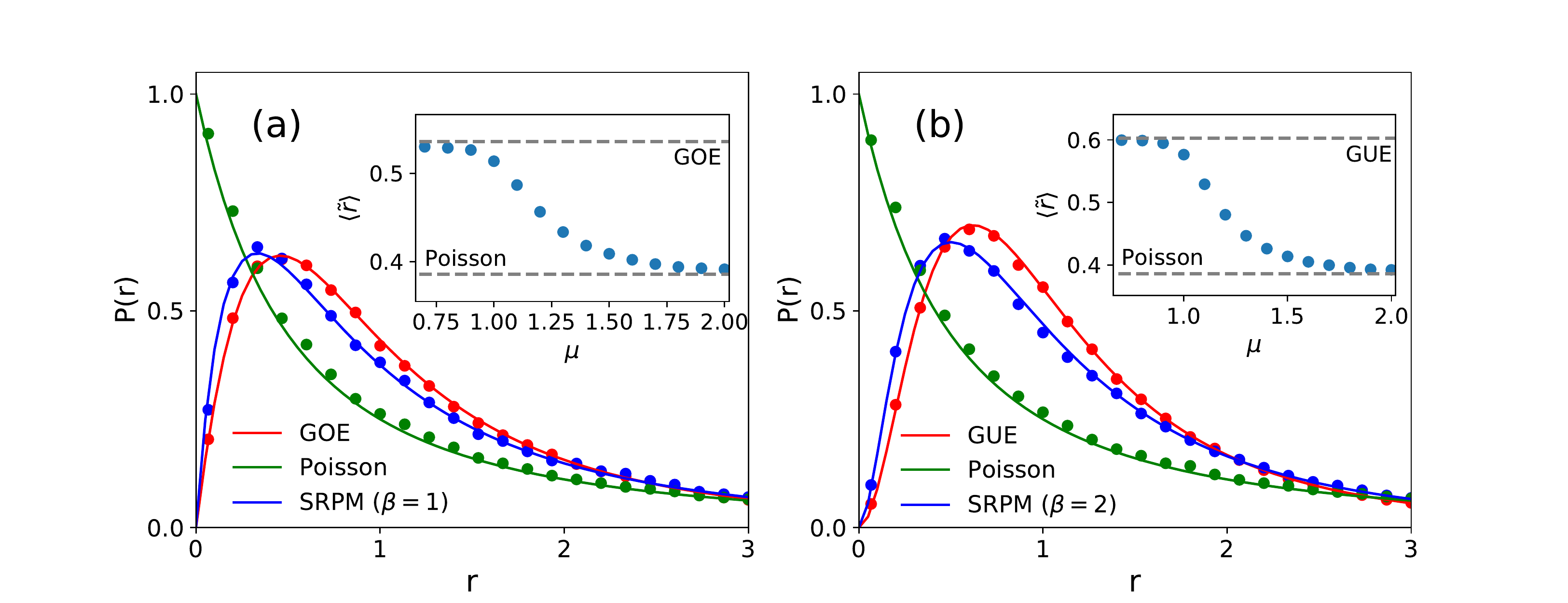}
\caption{$P(r)$ in (a) orthogonal and (b) unitary PRBM,
where the red,blue,green dots stand for $\mu=0.7,1.08,2$ in both sub-figures. Insets: Evolution
of $\langle\tilde{r}\rangle$ in respective cases.} \label{fig:demo}
\end{figure}

As the first demonstration, we plot $P\left( r^{\left( 1\right) }\right) $
of PRBM with varying $\mu $ in cases with $\beta =1,2$. Particularly, we aim
at three target distributions: WD for ergodic phase; Poisson for MBL phase;
and that of short-range plasma model (SRPM)\cite{SRPM} with semi-Poisson
distribution, which is commonly believed to be the critical distribution at
the ergodic-MBL transition point\cite{Serbyn,Sierant19}. The expressions of
them are\cite{Atas,Rao21,SRPM}%
\begin{equation}
P(r)=\left\{
\begin{array}{ll}
Z_{\beta }\frac{\left( r+r^{2}\right) ^{\beta }}{\left( 1+r+r^{2}\right)
^{1+3\beta /2}}\text{,} & \quad \text{WD;} \\[1mm]
Z_{\beta }^{^{\prime }}\frac{r^{\beta }}{\left( 1+r\right) ^{2\left( \beta
+1\right) }}\text{,} & \quad \text{SRPM;} \\[1mm]
\frac{1}{\left( 1+r\right) ^{2}}\text{,} & \quad \text{Poisson.}%
\end{array}%
\right.  \label{equ:nearest}
\end{equation}%
where $Z_{\beta }$ and $Z_{\beta }^{^{\prime }}$ are normalization factors.
The fittings results are in Fig.~\ref{fig:demo}, as can be seen, PRBM with $%
\mu =0.7,1.08,2$ accurately reproduce the three target distributions in both
orthogonal and unitary cases.

To further illustrate the level evolution, we compute the average value of a
variant spacing ratios\cite{Oganesyan}, which is $\widetilde{r}_{i}=\frac{%
\min \left\{ s_{i+1},s_{i}\right\} }{\max \left\{ s_{i+1},s_{i}\right\} }$. $%
\langle \widetilde{r}\rangle $ takes different values in each ensemble\cite%
{Atas}, namely $\langle \widetilde{r}\rangle _{\text{GOE}}=0.536$, $\langle
\widetilde{r}\rangle _{\text{GUE}}=0.603$ and $\langle \widetilde{r}\rangle
_{\text{Poisson}}=0.386$. As shown in the insets of Fig.~\ref{fig:demo},
PRBM faithfully describes the continuous interpolation between WD and
Poisson. As the next step, we use PRBM to fit the eigenvalue statistics of
physical models, with particular attention paid for long-range level
correlations.

\section{Fitting Eigenvalue Statistics}

\label{sec3}

We first consider the paradigmatic orthogonal spin model with ergodic-MBL
transition, that is, the anti-ferromagnetic Heisenberg model with random
external fields\cite{Alet}, the Hamiltonian reads%
\begin{equation}
H_{0}=\sum_{i=1}^{L}\mathbf{S}_{i}\cdot \mathbf{S}_{i+1}+h\sum_{i=1}^{L}%
\varepsilon _{i}S_{i}^{z}\text{,}  \label{equ:H}
\end{equation}%
The anti-ferromagnetic coupling strength is set to be $1$, and $\varepsilon
_{i}$s are random variables within range $\left[ -1,1\right] $. This
Hamiltonian commutes with $S_{T}^{z}=\sum_{i}S_{i}^{z}$ and allows us to
reach larger systems by focusing on one sector. In this work, we choose $%
L=16 $ in the $S_{T}^{z}=0$ sector, with Hilbert space dimension $%
C_{16}^{8}=12870 $. We exactly diagonalize $H_0$ at various random strengths
with $400$ samples taken at each point, and select $400$ eigenvalues in the
middle to determine $P\left( r^{\left( n\right) }\right) $.

\begin{figure*}[t]
\centering
\includegraphics[width=1.8\columnwidth]{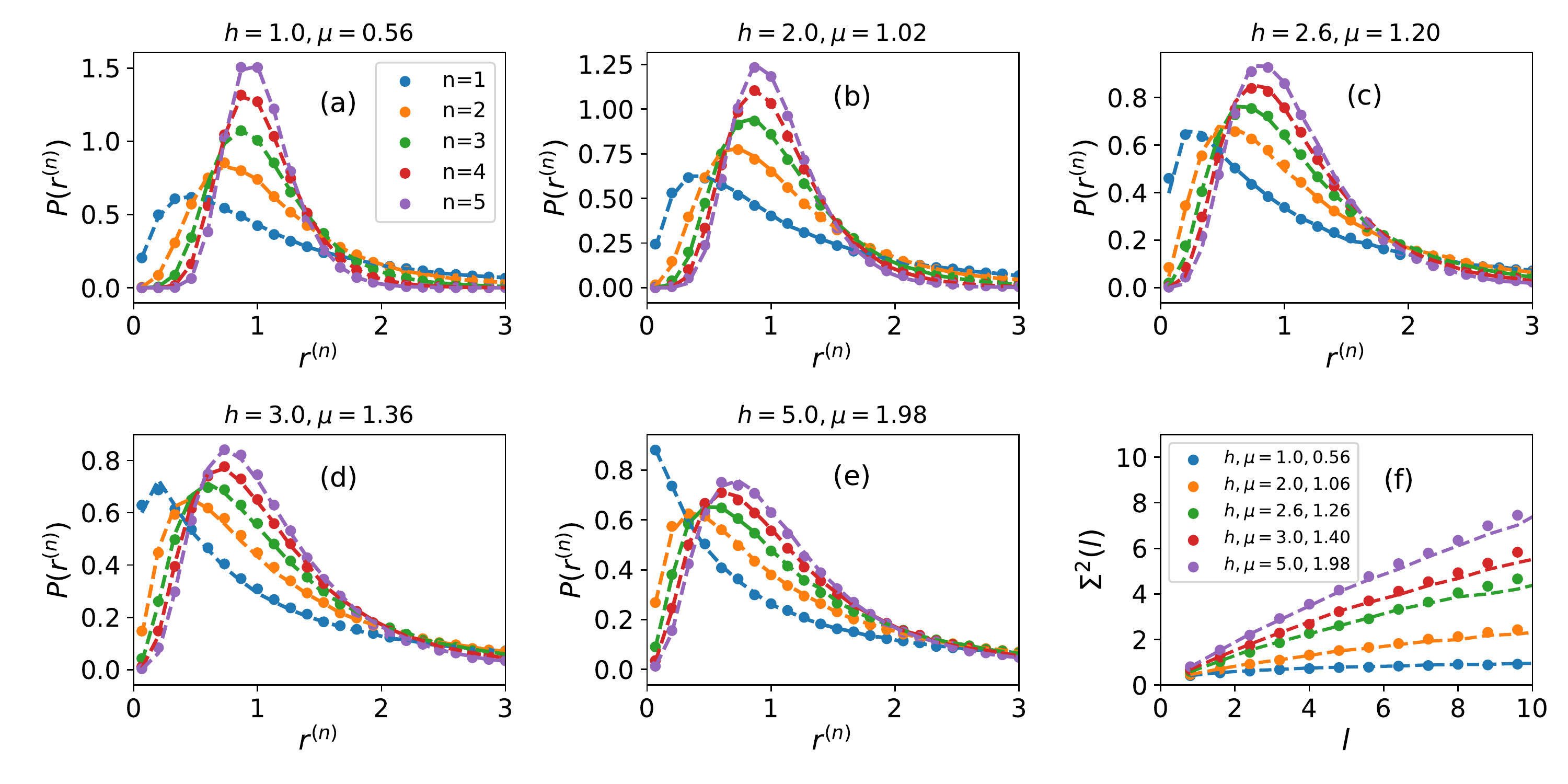}
\caption{(a)-(e) Comparisons of the spacing ratio distributions $P(r^{(n)})$
up to $n=5$ between the orthogonal spin model with various disorder
strengths and PRBM with $\protect\beta=1$, where the title of each
sub-figure indicates the disorder strength $h$ and fitted $\protect\mu$.
(a)-(e) share the same legend. (f) Evolution of number variance $\Sigma^2(l)$
of the physical model and PRBM, the fitted $\protect\mu$s are displayed in
the figure legend. The values of $\protect\mu$s fitted by $P(r^{(n)})$ and $%
\Sigma^2(l)$ match perfectly in the ergodic ($h=1$) and MBL ($h=5$) phase,
and their deviations in the transition region ($h=2\sim3$) are kept within $%
5\%$. Dots and lines stand for physical data and PRBM respectively in all
sub-figures.}
\label{fig:orth}
\end{figure*}

This physical model displays an ergodic-MBL evolution in the range $h\in
\left( 1,5\right) $, we select several representative points over the entire
phase diagram, and compared the resulting $P\left( r^{\left( n\right)
}\right) $ to those of PRBM with fitted parameter $\mu$, the results are
shown in Fig.~\ref{fig:orth}(a)-(e). Remarkably, we observe nearly perfect
agreements between PRBM and physical model in all the cases considered. For
clarity we only display the fitting results up to $n=5$ in Fig.~\ref%
{fig:orth}(a)-(e), while the agreements maintain up to much longer ranges
even in the transition region ($h\in \left( 2,3\right) $ in this system
length).

To further study the long-range level correlations, we employ the number
variance $\Sigma ^{2}\left( l\right) $\cite{Mehta,Haake}. Unlike spacing
ratios, $\Sigma ^{2}\left( l\right) $ is very sensitive to the concrete
unfolding strategy, and here we used the normal unfolding with third-order
polynomials. Fig.~\ref{fig:orth}(f) displays the results, where the optimal $%
\mu$s of PRBM are independently fitted through $\Sigma ^{2}\left( l\right) $%
. Again, we observe very good agreements. Notably, the optimal values of $%
\mu $ fitted by $P\left( r^{\left( n\right) }\right) $ and $\Sigma
^{2}\left( l\right) $ match perfectly for cases deep in the ergodic ($h=1$)
and MBL ($h=5$) phase, and the relative deviations in the transition region
are kept within $5\%$ -- a satisfying accuracy.

We stress here the fittings of high-order spacing ratios $P\left( r^{\left(
n\right) }\right) $ with $n>1$ and $\Sigma ^{2}\left( l\right) $\ are
necessary, since it is known that different random matrix models may have
very similar local level statistics\cite{Shukla,Shukla2}. For example, the
Brownian ensemble (linear combination of random matrix in WD and Poisson
ensemble) and the Gaussian $\beta $ ensemble are both capable of describing $%
P\left( r^{\left( n\right) }\right) $ with $n=1$ for the ergodic-MBL
transition, but their difference soon become non-negligible when
higher-order spacing ratios are considered\cite{Rao212}.

\begin{figure*}[t]
\centering
\includegraphics[width=1.8\columnwidth]{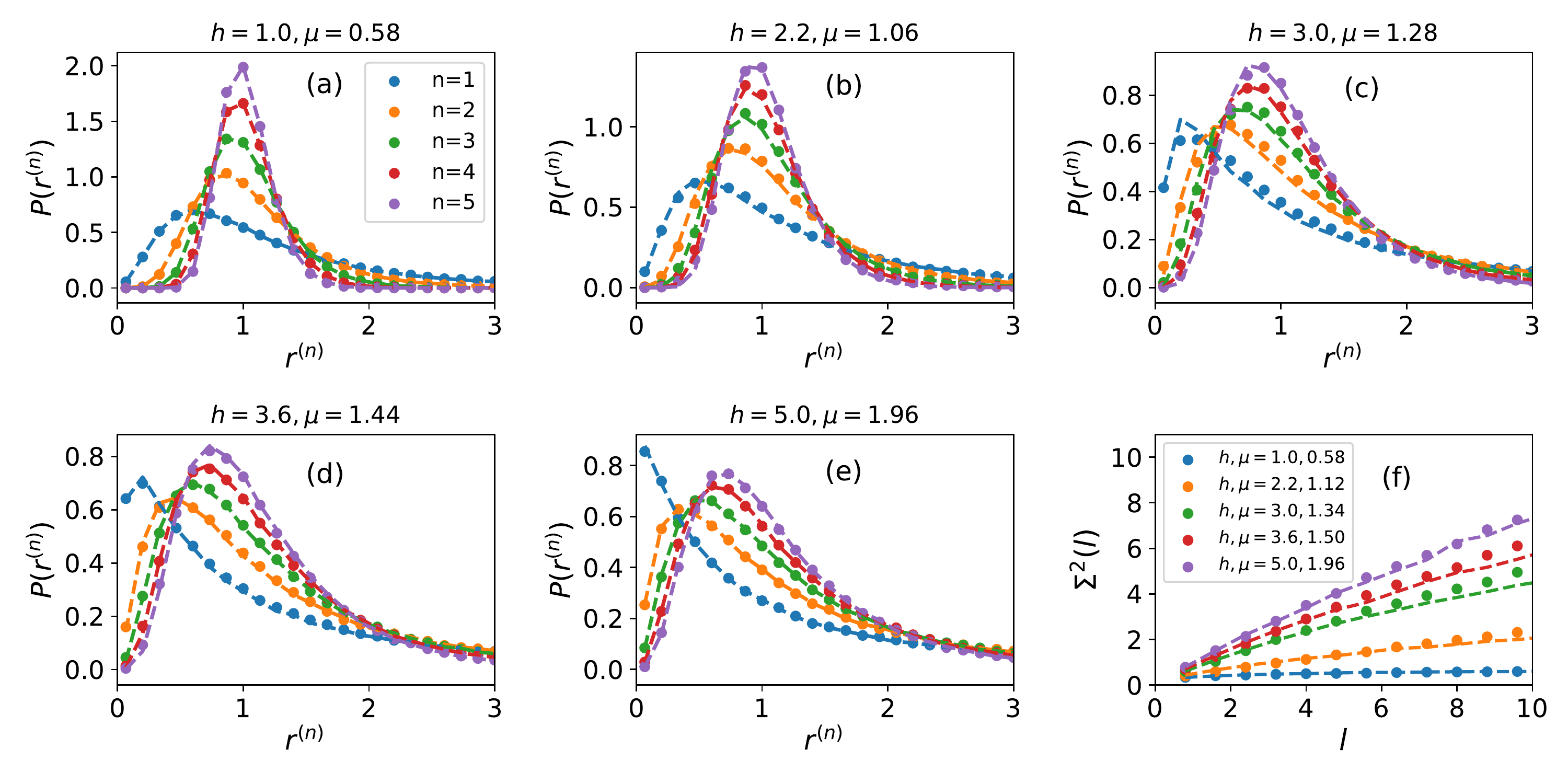}
\caption{(a)-(e) Spacing ratio distributions $P(r^{(n)})$ in the unitary
spin model without time-reversal symmetry and PRBM with $\protect\beta=2$,
the randomness strength $h$ and fitted $\protect\mu$ are displayed in the
title of each sub-figure. (f) Evolution of number variance $\Sigma^2(l)$ of
the physical model and PRBM with fitted $\protect\mu$s displayed in the
figure legend. Dots and lines stand for physical data and PRBM in all cases.}
\label{fig:unit}
\end{figure*}

To illustrate the power of PRBM in the unitary ($\beta =2$) case, we break
the time-reversal symmetry of $H_{0}$ in Eq.~(\ref{equ:H}) by adding a term $%
JH_{1}$, where\cite{Avishai2002}
\begin{equation}
H_{1}=\sum_{i=1}^{L-2}\mathbf{S}_{i}\cdot \left( \mathbf{S}_{i+1}\times
\mathbf{S}_{i+2}\right) \text{,}
\end{equation}%
and we fix $J=0.2$ without loss of generality. This model also preserve $%
S_{T}^{z}$, and we likewise study in the $S_{T}^{z}=0$ sector of an $L=16$
system, with the rest settings identical to the previous model. The fittings
of $P(r^{(n)})$ and $\Sigma ^{2}(l)$ by PRBM\ in various disorder strengths
are collected in Fig.~\ref{fig:unit}(a)-(e) and (f) respectively. Once
again, we observe nearly perfect agreements between PRBM and physical data
in all cases. The deviations are slightly larger than the orthogonal case in
the intermediate region, suggesting a larger finite size effect, this
situation also happen in models without $S_{T}^{z}$ conservation (see the
appendix).

It is worth mentioning that the representative randomness strengths are
deliberately selected. To be specific, the ergodic-MBL transition point in
all these spin systems have been studied in Ref.~[\onlinecite{Rao21}], and
we correspondingly select five points to represent the ``entire phase
diagram'', i.e. the cases deep in the ergodic/MBL phases, right at the
transition point, and intermediate between the ergodic/MBL phases and
transition point. We believe that such choices are sufficient to represent
the whole phase diagram.

Up to now, we have verified the validity of PRBM in modeling the eigenvalue
statistics of physical systems both with and without time-reversal symmetry.
We can further show this is not model-dependent by fitting physical models
without $S_{T}^{z}$ conservation, which is presented in the appendix. Next
we turn to the study of eigenvectors.

\section{Scaling of Entanglement Entropy}

\label{sec4}

The eigenvector of PRBM is a bit tricky since it lacks a direct
correspondence to physical observables. Nevertheless, we can artificially
view $H$ as the Hamiltonian matrix for a system of fictitious local degrees
of freedoms, then we are able to study the general quantities of
eigenvectors. Here we consider the quantity of modern interest, that is, the
entanglement entropy (EE). In this section we focus on the orthogonal PRBM
with $\beta =1$.

To get the EE of PRBM, we treat $H$ with dimension $N\times N$ as the
Hamiltonian describing the interaction between $L_{f}$ fictitious spin-$1/2$%
s, where $L_{f}=\log _{2}N$. A typical eigenvector $|\varphi \rangle $ is
therefore represented by an $N$-dimensional column vector. Then by dividing
the system into equal halves A and B with $L_{f}/2$ spins, $|\varphi \rangle
$ is decomposed as $|\varphi \rangle =\sum_{ij}\varphi _{ij}|\varphi
_{A}\rangle _{i}|\varphi _{B}\rangle _{j}$ where $|\varphi _{A}\rangle _{i}$
and $|\varphi _{B}\rangle _{j}$\ are both $\sqrt{N}$-dimensional column
vectors representing the subsystem's basis and $\varphi _{ij}$ is obtained
by reshaping the $N$-dimensional column vector $|\varphi \rangle $ into a $%
\sqrt{N}\times \sqrt{N}$ matrix $\varphi $. The reduced density matrix $\rho
_{A}=Tr_{B}\left( |\varphi \rangle \langle \varphi |\right) $ is thus equal
to $\varphi \varphi ^{\dagger }$ and EE is obtained through $S=-Tr_{A}\left(
\rho _{A}\log \rho _{A}\right) $.

\begin{figure}[t]
\centering
\includegraphics[width=0.8\columnwidth]{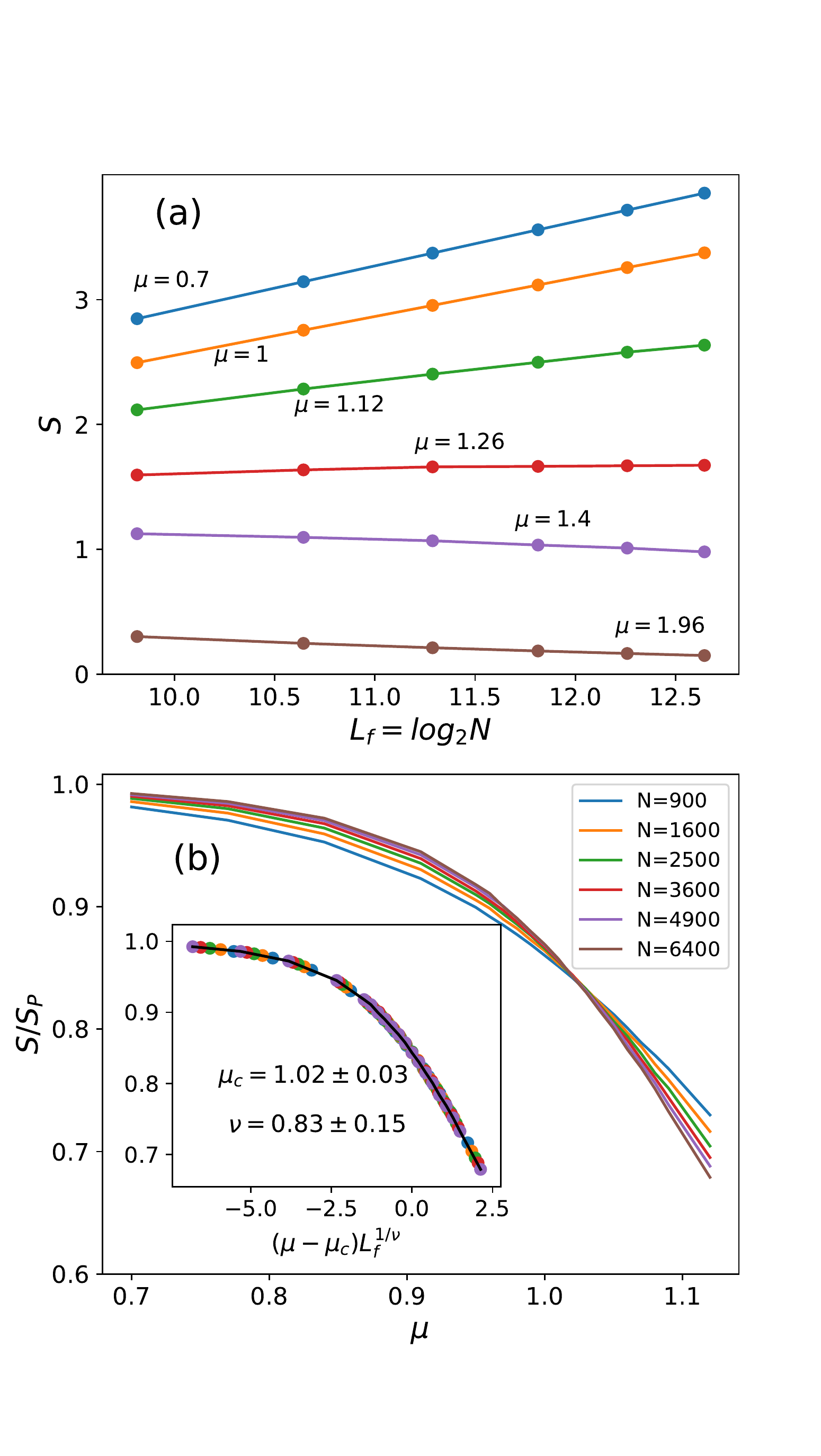}
\caption{(a) Average half-chain EE with respect to fictitious system length $L_f$ in PRBM with different $\mu$s.
(b) Evolution of EE $S$ divided by the Page value $S_P$,
a crossing indicates a critical point at $\mu\sim1$. For eye's convenience to observe the
crossing, data for larger $\mu$ are omitted. Inset: Finite-size collapse for $S/S_P$
near the critical region, the estimated critical point $\mu_c$ and critical exponent $\nu$ are displayed.}
\label{fig:EE}
\end{figure}

We compute the EE averaged among the middle $200$ eigenvectors of $H$ with
parameter $\mu $, and determine its scaling with the fictitious system
length $L_{f}$, the results are in Fig~\ref{fig:EE}(a). We see that for $\mu
$ in the ergodic phase, $S$ grows linearly with $L_{f}$, which reflects the
volume-law behavior; while for large $\mu $, EE saturates to a small value,
reflecting the area-law behavior in an one-dimensional MBL phase.

We then draw the evolution of $S$ divided by the Page value $S_{P}=0.5(L_{f}%
\text{log}(2)-1)$ for a pure random state\cite{Page} as a function of
parameter $\mu $, the results are shown in Fig~\ref{fig:EE}(b). As expected,
$S/S_{P}$ decreases from $1$ in ergodic phase to $0$ for MBL phase with
increasing $\mu $. The curves with different matrix dimensions $N$ cross at
the critical point $\mu _{c}\simeq 1$, confirming the criticality in this
case.

We further zoom in the transition region to determine the critical exponent
by assuming $S/S_{P}=g\left( \left( \mu -\mu _{c}\right) L_{f}^{1/\nu
}\right) $, with $g$ being an unspecified continuous function. The
finite-size collapes is shown in the inset of Fig.~\ref{fig:EE}(b), where we
find $\mu _{c}=1.02\pm 0.03$ and $\nu =0.83\pm 0.15$. This value of $\nu $
is smaller, while close to the one $\nu \sim 1$ found by exact
diagonalizations for the orthogonal spin model in Eq.~(\ref{equ:H})\cite{RD}
, which supports the validity of PRBM in modeling the eigenvectors of
physical system.

It's worth emphasizing that the maximal matrix dimension in Fig.~\ref{fig:EE}
is $N=6400$, which is much smaller than the spin model in Ref.~[%
\onlinecite{RD}] (up to $L=18$ with dim$(H)=48620$). This fact strongly
hints the universal properties of MBL\ transition is highly distilled in the
power-law construction of PRBM. This is of course far from being fully
understood, and further insights can be gained by constructing a physical
model whose Hamiltonian bears the form of PRBM, which is left for a future
study.

\section{Conclusion and Discussion}

\label{sec5}

In this work, we aim to search for an effective mathematical model that can
phenomenologically describe the ergodic-MBL transition in physical spin
models. We showed that the PRBM with fixed $B=1$ and single tuning parameter
$\mu $ is such a proper choice. For one side, we show that PRBM gives nearly
perfect description for the eigenvalue statistics -- both on short and long
ranges -- on the entire phase diagram with ergodic-MBL transition, in
systems both with and without time-reversal symmetry. Moreover, for the
eigenvectors, it's shown the EE of PRBM displays an evolution from
volume-law to area-law behavior which signatures an ergodic-MBL transition,
and the critical exponent is found to be $\nu =0.83\pm 0.15$, close to the
value obtained by exact diagonalization in physical model. To the best of
our knowledge, PRBM is the first single-parameter random matrix model that
models both the eigenvalues and eigenvectors of disordered physical systems.

It's remarkable that the PRBM with simple construction and single parameter
can model both the eigenvalues and eigenvectors of a physical system, which
suggests the universal, model-independent physics of MBL system is highly
distilled in PRBM. One way to interpret this result is to view PRBM as the
late-stage matrix during the renormalization flow that diagonalize the
Hamiltonian\cite{Flow,Flow1,Flow2,Flow3}, where the diagonal and
off-diagonal elements represent eigenergies of the local-integral-of-motions
(LIOMs) and the interactions between them. Actually, hints for such a
power-law interaction between LIOMs have already been noticed in Ref.~[%
\onlinecite{Flow3}]. Another approach is to view MBL as the single particle
Anderson localization in the Fock space\cite{Nayak,Fock}, where the diagonal
and off-diagonal elements stand for the Hatree-Fock and hopping terms. It is
an interesting question to construct a corresponding physical model in this
regard.

A noticeable property of PRBM is that it's dense, while in most physical
systems with finite interaction range the Hamiltonian matrix is highly
sparse. It's then natural to ask whether the sparsity should be taken into
account when constructing an effective random matrix model. Such a question
has been attempted in Ref.[\onlinecite{Papenbrock}], where the authors
constructed the \textquotedblleft scaffolding\textquotedblright\ random
matrix ensemble that is much sparser than PRBM for the metal-insulator
transition in interacting Fermionic system. It's an interesting question
whether the scarffolding random matrix can describe the spectral evolution
as accurately as PRBM does. On the other hand, given the efficiency of PRBM
in describing the unitary spin systems with next-nearest neighboring
interaction (Fig.~\ref{fig:unit}) and with only nearest neighboring
interaction (Fig.~\ref{fig:unit1} in the appendix), it's tempting to
conclude that the interaction range does not affect the level evolution.
However, we also note a recent work that studies the Haldane-Shastry model
with infinite interaction range, where the authors claimed the spectral
statistics is neither WD nor Poissonian even when the randomness is very
large\cite{HS}. Therefore, the effect of increasing interaction range
remains to be explored.

The most interesting results are the entanglement behaviors in Sec.\ref{sec4}%
. The results in Fig.~\ref{fig:EE}(a) indicate that PRBM contains an
area-law to volume-law transition that hints an ergodic-MBL transition,
while the physical correspondence of the auxiliary bipartition is still
unclear since PRBM has no direct sense of space or locality. In other words,
it's not clear whether such an auxiliary bipartition is more close to the
normal left-right bipartition or the \textquotedblleft comb
bipartition\textquotedblright\ (an extensive bipartition of system into
alternating subsystems)\cite{Khaymovich1,Khaymovich2}. Although the critical
exponent $\nu $ extracted in Fig.~\ref{fig:EE}(b) supports the former, a
detailed evaluation is possible only if we find a physical model whose
Hamiltonian bears the form of PRBM. We also do not discuss the existence of
weak ergodicity breaking of the PRBM in the range $0.5<\mu <1$\cite%
{Sieber,Khaymovich3,Khaymovich4}, which may not affect the spectral
statistics but have important influence on the eigenvector structures.

The construction of PRBM relies only on symmetry, regardless of the type of
randomness. Therefore, it's natural to ask whether it can describe the
Griffiths effect, which distinguishes the MBL induced by random disorder
from that by quasi-periodic potential. It is suggested\cite{RD,Sierant19}
the existence of Griffiths regime can be revealed by the peak of
sample-to-sample variance $V_{S}$ of the mean spacing ratio $\langle
\widetilde{r}\rangle $, which is, however, absent in the PRBM. A possible
way to induce the peak of $V_{S}$ is to construct a linear combination of
PRBMs with different $\mu$s\cite{Sierant19}, which unavoidably introduces
extra parameters. Therefore, a single-parameter random matrix model
accounting for the Griffiths regime remains to be explored.

We expect that our work paves the way to the understanding of MBL and
related systems in several ways. First, our model is the first
single-parameter random matrix model that reproduces both the short and long
range eigenvalue statistics of an MBL system, which makes the quantitative
description for the evolution between ergodic and MBL\ phase possible.
Second, there is an existing discrepancy on the value of critical exponent $%
\nu $ obtained by exact diagonalization and renormalization group\cite%
{RD,SXZhang}, it is then attracting to perform certain renormalization to
the PRBM, which may help to resolve this debate. Last but not least, we have
also computed the EE in the unitary PRBM, and we observe totally similar
scaling behaviors and critical exponent to the orthogonal PRBM, we note
similar results have been noted in Ref.[\onlinecite{PRBMS}]. This hints the
critical property of the unitary spin model is also similar to the
orthogonal model. Understanding the effect of system's symmetry on MBL
transition is certainly an important task, and we left it to a future study.

\section{Acknowledgements}

The author acknowledges M. Haque and I. Khaymovich for stimulating and
helpful discussions. This work is supported by the National Natural Science
Foundation of China through Grant No.11904069.

Data Availibility Statement: The data that support the figures within this
paper are available from the corresponding author upon reasonable request.

\appendix

\section{Fitting Spin Models without $S_{T}^{z}$ conservation}

In this section we use PRBM to fit the eigenvalue statistics of random spin
models without $S_{T}^{z}$ conservation, the Hamiltonian is as follows
\begin{equation}
H=\sum_{i=1}^{L}\mathbf{S}_{i}\cdot \mathbf{S}_{i+1}+\sum_{i=1}^{L}\sum_{%
\alpha =x,y,z}h^{\alpha }\varepsilon _{i}^{\alpha }S_{i}^{\alpha }\text{.}
\end{equation}%
We will also consider the cases both with and without time-reversal
symmetry. For the former, it is the case with $h^{x}=h^{z}=h\neq 0$ and $%
h^{y}=0$; while for the latter, it is $h^{x}=h^{y}=h^{z}=h\neq 0$. For both
models we simulate an $L=13$ system, with Hilbert space dimension $%
2^{13}=8192$. In all cases, the number of eigenvalue spectrum samples is $%
400 $, and we take $400$ eigenvalues in the middle to determine $P\left(
r^{\left( n\right) }\right) $.

\begin{figure*}[h]
\centering
\includegraphics[width=1.8\columnwidth]{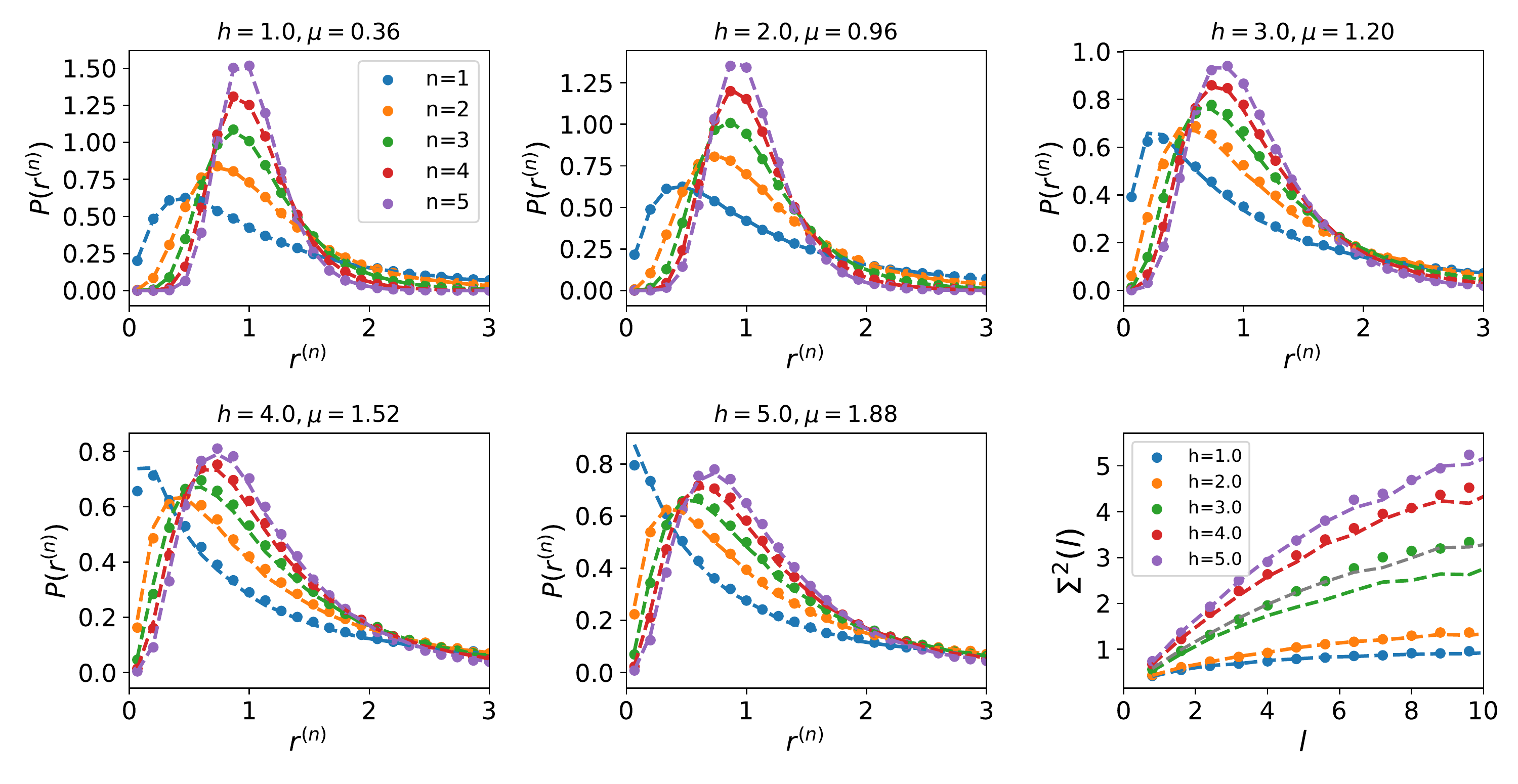}
\caption{The distributions of $P(r^{(n)})$ in the orthogonal spin model
without $S_T^z$ conservation with comparisons to PRBM, the title of each
sub-figure indicates the randomness strength $h$ and fitted $\protect\mu$.
Bottom right: $\Sigma^2(l)$ of the physical model and PRBM with $\protect\mu$
fitted through $P(r^{(n)})$, they fit quite well except for the transition
region $h\sim3$. Grey dashed line: $\Sigma^2(l)$ of PRBM at $\protect\mu%
=1.26 $. Dots and lines stand for physical data and PRBM in all cases.}
\label{fig:orth1}
\end{figure*}

For the orthogonal case, we likewise take several representative randomness
strengths to determine spacing ratio distributions $P\left( r^{\left(
n\right) }\right) $ and number variance $\Sigma ^{2}\left( l\right) $, and
compare them to those of orthogonal PRBM with $\beta=1$, the results are
collected in Fig.~\ref{fig:orth1}. This model suffers less from finite-size
effect, hence the fittings are close to perfect. Moreover, the optimal
parameter $\mu$ fitted by $P\left( r^{\left( n\right) }\right) $ and $\Sigma
^{2}\left( l\right) $ are very close in most of the phase diagram, except
for the transition region $h\sim3$, as can be viewed from the last
sub-figure of Fig.~\ref{fig:orth1}.

\begin{figure*}[t]
\centering\includegraphics[width=1.8\columnwidth]{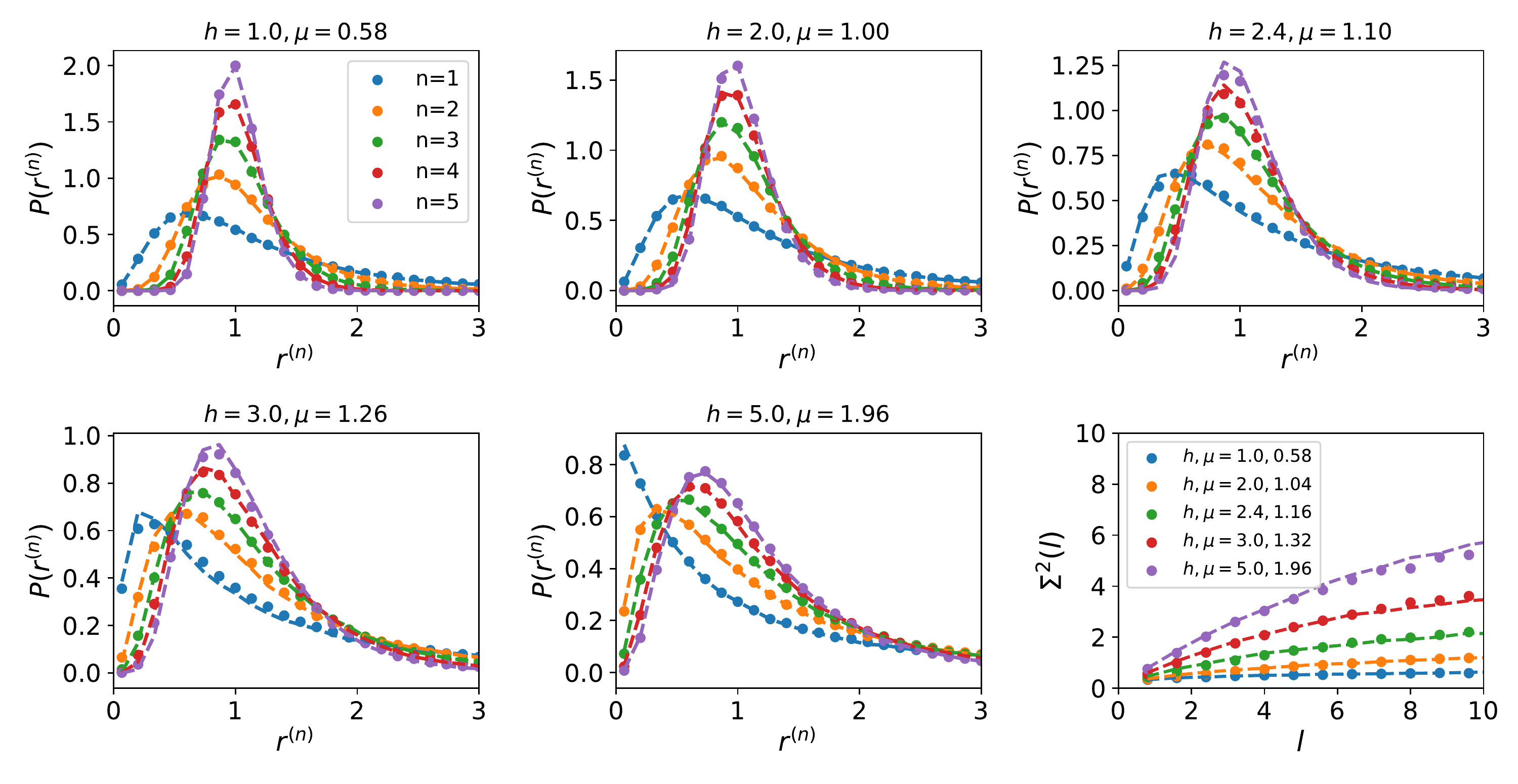}
\caption{The distributions of $P(r^{(n)})$ in the unitary spin model without
$S_T^z$ conservation at various disorder strengths with comparisons to PRBM
with $\protect\beta=2$ and fitted $\protect\mu$. Bottom right: $\Sigma^2(l)$
of physical model and PRBM. Note here the values of $\protect\mu$ are fitted
by $\Sigma^2(l)$ as listed in the figure legend, they have minor deviations
from those fitted through $P(r^{(n)})$, but the relative errors are
controlled within 5\%. Dots and lines stand for physical data and PRBM in
all cases.}
\label{fig:unit1}
\end{figure*}

For the time-reversal breaking case with $h^{x}=h^{y}=h^{z}=h$, the fitting
results are displayed in Fig.~\ref{fig:unit1}. The finite size effects are
slightly larger than the orthogonal case, just like in the models with $%
S_T^z $ conservation as discussed in the main text. Nevertheless, the
deviations between the values of $\mu$ fitted by $P\left( r^{\left( n\right)
}\right) $ and $\Sigma ^{2}\left( l\right) $ are still less than 5\%.

\end{document}